\documentclass[authoryear,12pt,3p,letter]{jowarticle}
\usepackage{graphicx}
\usepackage{amsthm,amsmath}
\usepackage{amssymb}
\usepackage{amsfonts}
\usepackage{natbib}
\usepackage{setspace}
\usepackage[all]{xy}
\usepackage{enumitem}
\usepackage{titlesec}
\usepackage{mathrsfs}
\makeatletter
\renewcommand\section{\@startsection{section}{1}{\z@}{-3.25ex plus -1ex minus -.2ex}{1.5ex plus .2ex}{\normalsize\bf}}
\renewcommand\subsection{\@startsection{subsection}{2}{\z@}{-3.25ex plus -1ex minus -.2ex}{1.5ex plus .2ex}{\normalsize\bf}}
\renewcommand\subsubsection{\@startsection{subsubsection}{3}{\z@}{-3.25ex plus -1ex minus -.2ex}{1.5ex plus .2ex}{\normalsize\bf}}
\makeatother

\newtheorem{innercustomgeneric}{\customgenericname}
\providecommand{\customgenericname}{}
\newcommand{\newcustomtheorem}[2]{%
  \newenvironment{#1}[1]
  {%
   \renewcommand\customgenericname{#2}%
   \renewcommand\theinnercustomgeneric{##1}%
   \innercustomgeneric
  }
  {\endinnercustomgeneric}
}

\newcustomtheorem{prop*}{Proposition}
\newcustomtheorem{cor*}{Corollary}

\begin{document}
\begin{frontmatter}
\title{Some Philosophical Prehistory of the (Earman-Norton) Hole Argument}
\author{James Owen Weatherall}\ead{weatherj@uci.edu}
\address{Department of Logic and Philosophy of Science\\ University of California, Irvine}
\begin{abstract}
The \emph{celu} of the philosophical literature on the hole argument is the 1987 paper by Earman \& Norton [``What Price Space-time Substantivalism?  The Hole Story'' \emph{Br. J. Phil. Sci}].  This paper has a well-known back-story, concerning work by Stachel and Norton on Einstein's thinking in the years 1913-15.  Less well-known is a connection between the hole argument and Earman's work on Leibniz in the 1970s and 1980s, which in turn can be traced to an argument first presented in 1975 by Howard Stein.  Remarkably, this thread originates with a misattribution: the argument Earman attributes to Stein, which ultimately morphs into the hole argument, was not the argument Stein gave.  The present paper explores this episode and presents some reflections on how it bears on the subsequent literature.
\end{abstract}
\end{frontmatter}
\doublespacing
\section{Introduction}\label{sec:intro}

The hole argument was originally articulated by Einstein, in 1913, as a challenge to any ``generally covariant'' theory of gravitation. Briefly, it shows a certain sense in which general relativity---or really, any theory of geometric objects formulated on a manifold---might be said to be ``indeterministic''.\footnote{I do not believe the hole argument shows what it is sometimes alleged to show (see \citet{WeatherallHoleArg} and \citet{Bradley+etal}).  But as I discuss below, there is an important mathematical (and conceptual) issue in the vicinity of the hole argument, regarding what it means for a system of differential equations to admit an initial value formulation ``up to diffeomorphism''.  See, for instance, \citet{GerochGauge} for a discussion of this issue.}  But by the end of 1915, Einstein had rejected the hole argument, in part because he had developed what he believed to be a convincing counter-argument, often referred to as his ``point-coincidence argument.''.\footnote{Einstein first published general relativity in November 1915, whereas the first known statement of the point coincidence argument came in a letter in December 1915, to Ehrenfest \citep[see, e.g.,][]{GiovanelliKretschmann}.  So it is not perfectly clear whether he first rejected the hole argument on the grounds of the point coincidence argument, and only then proceeded to discover Einstein's equation; or if he returned to ``general covariance'' for other reasons, and then, upon finding his eponymous equation, found an argument to support it against hole-based challenges.  I am grateful to an anonymous referee for pressing this point.}

For the next 70 years, the hole argument received little attention.\footnote{This gloss is, perhaps, unfair: Marco Giovanelli, in unpublished work, has argued that the hole argument, and the point-coincidence argument, were rediscovered multiple times during the ensuing decades, by groups working on various approaches to quantum gravity.  But it was not widely discussed in the physics or philosophy of physics literatures.}  During the 1980s, however, it resurfaced: in their highly influential paper, ``What Price Space-Time Substantivalism?'', \citet{Earman+Norton} re-cast the hole argument as an argument that a certain variety of substantivalism---here understood as the view that ``points'' of space and time exist independently of and ontologically prior to the events that occur at those points---is committed to a pernicious form of  indeterminism.  They went on to argue that one can avoid this variety of indeterminism by accepting what they call ``Leibniz equivalence'', which they take to be a hallmark of relationism.  In the subsequent three decades, many philosophers, attracted to the form of substantivalism Earman and Norton reject, have offered responses to the hole argument, leading to a now-large literature.\footnote{For recent reviews of this literature, see \citet{NortonSEP}, \citet{PooleyOxford}, and \citet{StachelReview}.}

Reviews of the hole argument literature often mention that \citep{Earman+Norton} had a pre-history.  In 1980, John Stachel presented a paper in Jena on Einstein's search for general relativity, later published as \citep{Stachel}, in which he argued that the hole argument had lasting conceptual significance.  Similarly, John \citet{Norton1984} argued, on the basis of Einstein's Z\"urich notebooks, that the hole argument had played an essential role in Einstein's thinking about general covariance.  From this perspective, \citet{Earman+Norton} took insights from the historical literature on the hole argument and exposed their philosophical importance.

Less widely recognized, apparently, is that John Earman also discussed the hole argument prior to his 1987 paper with Norton.  For instance, it appears in several 1986 publications \citep{EarmanPD,Earman1986}.\footnote{\label{pubHistory1} In these 1986 publications, Earman cites the paper with Norton as forthcoming, suggesting that there was a delay before publication of the Earman-Norton paper.  That said, the published version of the 1987 Earman and Norton article reports that it was ``Received December 1986''; it is difficult to square this timing with Earman's citations in 1986, unless the December 1986 date reflects the receipt of the final version of the article, after it was accepted.  See also note \ref{pubHistory2}.}  Importantly, in \citep{Earman1986}, he explicitly connects the hole argument with a different argument, which he first presented in \citep{EarmanLSTLA} but also mentioned in \citep{EarmanPRM} and \citep{Earman1979}.  This other argument, on its face, has nothing to do with general relativity.  It concerns Leibniz.\footnote{See also \citep{EarmanWEST}, where he also draws connections between the hole argument and these earlier arguments about Leibniz. One of the central proposals of these papers is that one can recover a version of Leibnizian relationism by adopting an ``algebraic'' approach to space-time, both classically and in general relativity; this proposal has subsequently been discussed by \citet{Rynasiewicz1992}, \citet{Bain}, and \citet{Rosenstock+etal}.}

This thread represents a distinct and important part of the pre-history of the hole argument paper. It is in this work that Earman develops a line of argument connecting indeterminism to the idea that certain formulations of physical theories might have ``excess structure''.  And perhaps most remarkably of all, considering the lasting influence of the argument, is that the thread appears to originate in a misattribution.  In particular, Earman attributes the argument concerning Leibniz that he discusses to a talk by Howard Stein, first delivered in Minnesota in 1975 and later published as \citep{SteinPrehistory}.  But a careful reading of Stein's paper, and a comparison with Earman's recovery of the argument, reveals that Earman's version is importantly different from Stein's.  My goal in the present paper is to draw out these differences and show why they matter.  In the end, I will argue, they reflect two different ways of understanding how space-time theories represent the physical world---with different consequences for what significance we should take the hole argument to have.\footnote{To be clear: I do not emphasize this misattribution to criticize or embarrass Earman.  Rather, it is because Earman's understanding of these issues has become so influential in the subsequent literature, and such a core part of the tradition in philosophy of physics, that it can be difficult to imagine other ways of thinking about them.  For this reason, I take it to be important that when Earman began working on these issues, there was already a different perspective available; and that, possibly because Earman himself attributed a subtly different argument to Stein from the one Stein apparently intended, those two perspectives have not been carefully distinguished in the subsequent literature.  I should also emphasize that the fact that Earman's version of the argument differs from Stein's does not, by itself, reveal a shortcoming of Earman's argument (or Stein's)---after all, both versions might be compelling.  But it does show that, even at its origins, there have been disagreements concerning what significance the hole argument has---and in particular, about whether the standard formalism of general relativity somehow suggests or implies ``substantivalism''.}

The remainder of this paper will be organized as follows.  First I will present Earman's work on Leibniz and explain its connection, in his writings, to the hole argument.  In the following section, I will introduce Stein's version of the argument and argue that it is importantly different from Earman's.  I will then make a first pass at distinguishing the arguments, by describing two senses of ``determinism'' that seem to be operative in the two versions of the argument.  I then suggest that something deeper is going on, regarding how Earman understands the significance of manifolds in models of space-time theories.  I will conclude by bringing this discussion back to the hole argument.

\section{An alternative prehistory}\label{sec:Earman}

The hole argument may be stated as follows:\footnote{I do not review the basic formalism of general relativity here; for background, see \citet{Wald} or \citet{MalamentGR}, both of whom use essentially the same notation I do (including the so-called ``abstract index'' notation).} Let $(M,g_{ab})$ be a relativistic space-time---that is, a smooth manifold $M$ equipped with a smooth, Lorentz-signature metric, $g_{ab}$---and let $O\subsetneq M$ be the interior of a compact set of $M$.  Let $\varphi:M\rightarrow M$ be an automorphism of $M$ that acts as the identity outside of $O$, and which is not the identity on $O$. Then $(M,\varphi^*(g_{ab}))$ is also a relativistic space-time. The space-times $(M,g_{ab})$ and $(M,\varphi^*(g_{ab}))$ agree at all points $p\in M/O$, but they disagree in $O$. Thus no specification of $g_{ab}$ (or its derivatives) on $M/O$ could be sufficient to determine the values of $g_{ab}$ within $O$.

In a 1989 essay, ``Leibniz and the Absolute vs. Relational Dispute'', published in a collection of essays on Leibniz, Earman presents what is essentially this argument and then concludes ``since the two models are identical for all past times [i.e., ``before'' $O$] but differ in the future [i.e., in $O$], there is an apparent violation of Laplacian determinism'' \citep[p. 17]{Earman1989}.\footnote{\label{pubHistory2} Of course, 1989 comes after 1987, and so this article hardly counts as ``pre-history'' to the 1987 paper.  But I begin with this essay because (a) the statement of the relationship between the relativistic and Leibnizian arguments is particularly clear and (b) it demonstrates how Earman was thinking about the relationship between the two arguments around the time that he wrote the paper with Norton.  Indeed, although this chapter only appeared in print in 1989, there is good reason to believe it was written well before that---and, in fact, before 1987, because \citep{Earman+Norton} is cited in the chapter as ``Forthcoming'' in \emph{British Journal for Philosophy of Science}, with anticipated publication date of 1986. (Recall note \ref{pubHistory1}.) So this article should probably be seen as contemporaneous with \citep{Earman+Norton}.  I will presently trace this connection back into his earlier work.}  In this regard, the treatment is much the same as in \citet{Earman+Norton}---and indeed, after stating the argument and drawing this conclusion, Earman cites that paper (as forthcoming).  But I want to draw attention to how Earman prefaces these remarks: he writes, ``The conflict [between space-time substantivalism and the possibility of determinism] can be brought out by a variant on the construction used above in Sec. III'' (p. 17).

The argument in Section III of that paper has nothing to do with general relativity, or with Einstein.  Instead, it is exclusively concerned with the structure of classical (i.e., non-relativistic) space-time theories.  It arises in connection with a range of now-familiar possible structures that one might attribute to space and time in the context of classical physics.\footnote{For further detail on these different classical space-time structures, see \citet{EarmanWEST} or \citet{WeatherallCST}.}  In all of these, one begins with a smooth, four-dimensional manifold, assumed to be diffeomorphic to $\mathbb{R}^4$, and one adds further structure, resulting in a hierarchy of ever-more-structured candidate space-times.  In the present discussion, Earman considers first (1) adding just ``simultaneity structure'', which amounts to taking $M$ to be (the total space of) a (trivial) fiber bundle over $\mathbb{R}$, i.e., to fixing an identification of $M$ with $\mathbb{R}^3\times\mathbb{R}$.  He then considers (2) endowing each of the ``leaves'' of this foliated structure with a Euclidean metric, $h^{ab}$, representing distances between simultaneous points (or events) in space.\footnote{I have written $h^{ab}$ with raised indices because I am understanding it as a field on $M$ that induces a Euclidean metric on each leaf of the foliation.  See \citet[\S 4.1]{MalamentGR} for discussion of this field, and of $t_a$, introduced presently.}  He then adds (3) a temporal metric (i.e., a metric $t_a$ on $\mathbb{R}$ in the foliation $\mathbb{R}^3\times \mathbb{R}$) to construct ``Leibnizian space-time''; (4) a standard of rotation to get ``Maxwellian space-time'';\footnote{He does not define a ``standard of rotation'' here, though he invokes the concept; in \citet{EarmanWEST}, he defines a standard of rotation as an equivalence class of covariant derivative operators agreement on the ``twist'' of certain vector fields.  An alternative, equivalent but in some sense more intrinsic, definition of a standard of rotation in this context is given by \citet{WeatherallComment}; see also \citet{SaundersRNP} for yet another equivalent way of defining a standard of rotation.} (5) a standard of ``straightness'' for curves (i.e., a covariant derivative operator) that are not tangent to leaves of the foliation to get ``Galilean space-time'' (or ``Neo-Newtonian space-time''); and so on.\footnote{He does not use any of these names in the paper in question, but since he introduced them himself in \citet{EarmanWEST}, it seems to me not too anachronistic, and much simpler, to just refer to these structures in the now-standard ways.  He goes on to introduce (6) ``Newtonian space-time'', which includes a standard of rest; (7) ``Aristotelian space-time'', which includes a privileged location in space; and (8) an unnamed structure wherein one has a privileged origin in time (and space).}

After introducing these structures, Earman writes, ``An interesting a priori argument in favor of a space-time structure at least as rich as that of 5. [Galilean space-time] was given by Howard Stein (1977)'' (p. 13).  He then states the argument as follows:
\begin{quote}\singlespacing Grant that the space-time should be rich enough to allow for the possibility that particle motions are deterministic. Then the space-times 1.--4. are ruled out. For whatever the laws of particle motion are, a space-time symmetry should be a symmetry of the laws, i.e., should carry a set of world lines that solve the equations of motion to another set that is also a solution. But the symmetries in 1.--4. ... have the property that a symmetry map $d$ can be chosen with the property that it is the identity for all $t\leq0$ but non-identity for $t> 0$. Applying this map $d$ to a solution to the laws of motion produces another solution that agrees with the first for all past times but disagrees in the future, a violation of Laplacian determinism. (p. 13)
\end{quote}
I will have much more to say about this argument in the sequel, but for now observe that it shares with the hole argument the feature that one starts with a solution of some (unstated, in this case) equations and then one acts with a certain map $d$ to produce a ``new'' solution that agrees with the old one in one region of space-time, but not everywhere.

Earman discusses the hole argument alongside Stein's ``interesting a priori argument'' in other places as well (though not in \citep{Earman+Norton}).  For instance, in his classic book \emph{A Primer on Determinism}, Earman devotes two sections (\S\S III.2-3) to Stein's argument and Leibniz's imagined response, and then later, in his discussion of the Cauchy problem in General relativity, he presents the hole argument and explicitly compares it to Stein's argument as presented in \S III.2 of that book. He also directly connects the two arguments in his 1989 book \emph{World Enough and Space-Time}, presenting Stein's argument on pp. 55--7, and then comparing it to the hole argument on p. 179.  And in his 1986 paper ``Why Space is not a Substance (At Least Not to First Degree),'' though he does not explicitly compare the two arguments, he presents Stein's argument (pp. 232-3), and then on the immediately following page, he presents the hole argument, and identifies the ``relativity physicists'' response (viz. to take isometric space-times to represent the same physical situations) with what he supposes Leibniz's response to Stein's argument would have been.

Given the subsequent attention devoted to the hole argument, one might imagine that it was somehow primary---that is, that Earman considered Stein's argument in order to better explain the significance of the hole argument, in which case one might think that he is reading Stein's argument through that lens.  Reading \emph{A Primer on Determinism} or \emph{World Enough and Space-Time} can certainly give that impression, insofar as the classical issues are presented as a \emph{prolegomenon} to understanding modern physics.  But there are reasons to think the opposite is true, at least as a matter of the evolution of Earman's thought.  In fact, Earman first discusses Stein's argument almost a decade earlier (and shortly after Stein first presented it), in his 1977 book chapter, ``Leibnizian Space-Times and Leibnizian Algebras''.  There he writes:
\begin{quote} \singlespacing A third difficulty was raised by Stein (1975) [sic]. If the history of a particle is represented by a timelike world line (i.e. a world line which is everywhere oblique to the planes of simultaneity) on [Leibnizian space-time], then determinism cannot hold. For among the automorphisms [of Leibnizian space-time] are those which are the identity on the portion of [Leibnizian space-time] on or below some given time slice but which differ from the identity above; such a mapping leaves fixed the entire past history of the particles while changing their future behavior. Since the automorphisms of the space-time time [sic] should be symmetries of the dynamical laws (whatever they are), there will be two solutions which describe the same particle histories for all past times but which describe different future behaviors. If one desires determinism, as Leibniz did, then one must either modify [Leibnizian space-time] or one must find a different way of describing the history of a particle. Both of these alternatives will be explored below. Ultimately, a combination of the two will be offered as an explication of Leibniz.\footnote{Throughout this passage, Earman refers to Leibnizian space-time as ``$L_1$'' in anticipation of defining other possible space-time structures.} \citep[p. 96]{EarmanLSTLA}
\end{quote}

Earman goes on to propose, apparently for the first time, that one can partially avoid this kind of indeterminism by moving to a ``Leibniz algebra'' \citep[\S 5]{EarmanLSTLA}.\footnote{Earman refers back to this argument---without citing Stein---in several other papers in the late 1970s, including \citep[p. 225]{EarmanPRM} and \citet[p. 272]{EarmanWLR}.}  A \emph{Leibniz algebra} is defined in analogy with Einstein algebras, as introduced by \citet{GerochEA}.  Briefly, given a manifold, one can define a certain algebra $C^{\infty}(M)$ of smooth scalar fields on that manifold, and then proceed to define all of the structures we usually associate with general relativity, such as a Lorentz signature metric or various fields representing matter, directly using this algebra of functions.  Geroch's proposal is that one could just as well begin with a suitable algebraic structure, without ever mentioning a manifold, and proceed from there.\footnote{See \citet{Nestruev} and \citet{Rosenstock+etal} for further details.}  As I discuss below, Earman's idea is that by avoiding mention of the manifold, one can avoid any implicit ``substantivalist'' commitments.

The important point for the present discussion, however, is that the Hole Argument is not mentioned in this paper; in 1977, it remained in the waste bin of history.  Earman was primarily motivated, it seems, by Leibniz exegesis---and to develop (and partially revise) his own views as expressed in his classic ``Who's Afraid of Absolute Space?'' \citep{EarmanWAAS}, concerning the history of arguments for relational characterizations of space and time.  This strongly suggests, perhaps even demonstrates, that Earman's thinking about the hole argument was influenced by his understanding of Stein's argument, which he had already written and reflected on in some detail before ever discussing (or, perhaps, encountering), the hole argument.

\section{Probing the source}\label{sec:Stein}

We saw in the last section that in the years immediately before and after the publication of \citep{Earman+Norton}, Earman described the hole argument as a relativistic version of an argument concerning Leibnizian space-time that he attributed to Stein.  We also saw that Earman had discussed this Steinian argument as early as 1977, in the context of an argument that to recover Leibniz's views would require a move to Leibniz algebras, or something like them.  But as I will presently argue, Earman's understanding of Stein's argument, and its significance, was importantly different from Stein's own---and indeed, on Stein's understanding, the argument does not support the conclusions Earman drew from it.

The argument Earman attributes to Stein originates in a paper given at a 1975 conference at the University of Minnesota, which Earman attended; Stein's paper was later published in the 1977 proceedings of that conference, as ``Some Philosophical Prehistory of General Relativity'' \citep{SteinPrehistory}.  Stein writes, of Leibnizian space-time:
\begin{quote}\singlespacing
Now consider any smooth map of space-time onto itself that preserves simultaneity and ratio of time intervals, and that restricts, on each instantaneous space, to a Euclidean automorphism.  Such a map is an automorphism of the entire Leibnizian structure, and should therefore be a symmetry of the dynamics: i.e., should carry dynamically possible systems of world-lines to dynamically possible systems of world-lines. This immediately leads to the crucial difficulty: take any time-interval $[t_1, t_2]$, and any time $t$ outside this interval; then there exists a mapping of the sort described that is the identity map within $[t_1, t_2]$ but not at $t$. It follows that the dynamics cannot be such
as to \emph{determine systems of world-lines on the basis of initial data} (for the automorphism just characterized preserves all data during a whole time interval---which may even be supposed to be infinite in one direction, say to include ``the whole past''---but changes world-lines outside that interval). \citep[p. 6, emphasis original]{SteinPrehistory}
\end{quote}

Stein's argument certainly bears a strong resemblance to Earman's.  In both cases, one considers an automorphism of Leibnizian space-time that acts as the identity before some time $t$, but is not the identity after that; and one argues in each case that, therefore, there is a certain sense in which the future is not determined by the past.  But there is an important, if subtle, difference.  Whereas on Earman's reconstruction, what has been established is that there exist ``solutions'' to any suitable Leibnizian laws of motion that agree up to a certain time and then disagree after that time---that is, that Laplacian determinism fails---on Stein's version, all that is established is that Leibnizian laws could not determine world-lines of particles from initial data.  The gap between these conclusions is that on Stein's version, one has ruled out deterministic laws that involve representations of motion of a certain sort; whereas on Earman's version, one has jumped to the stronger conclusion that there could be \emph{no} suitably deterministic Leibnizian dynamics.  This stronger inference apparently depends on a suppressed premise, which is that any conceivable deterministic dynamics for particle motion would necessarily determine world-lines for particles.

That Stein wished to reject this suppressed premise, and to deny the stronger inference that Earman draws, is made explicit in the passage that immediately follows the one I have already quoted.  Stein continues,
\begin{quote}\singlespacing It must not be thought that this argument demonstrates the impossibility of a deterministic Leibnizian dynamics; the situation is, rather, that all of the systems of world-lines that arise from one another by automorphisms have to be regarded as objectively equivalent (i.e., as representing what is physically one and the same actual history). But the argument does show that a Leibnizian dynamics cannot take the form of a system of ``differential equations of motion,'' for such equations precisely do determine the world-lines from initial data. Or to put what is essentially the same point in a more sophisticated way: in Leibnizian space-time the ``phase,'' or instantaneous state of motion, of a system of particles cannot be represented by an assignment of 4-vectors to the world-points of the particles at that instant. In short, the basic conceptual apparatus for a cogent formulation of a dynamics satisfying this version of ``Leibnizian relativity'' would have to be significantly different from the structural framework we are used to. So far as I know, the appropriate concepts have never been defined. (p. 6)\end{quote}

In other words, Stein explicitly rejects the claim that his argument rules out a deterministic Leibnizian dynamics, or that a theory set in Leibnizian space-time is incompatible with determinism---whereas as we have seen, it is precisely this conclusion that Earman wishes to draw from his version of Stein's argument.\footnote{Stein has reiterated this interpretation of his argument in recent correspondence, writing ``I meant to disavow any claim of a demonstrative refutation of the possibility of a deterministic theory in the context of such a space-time structure, but to point out that such a theory could not have the form we now regard as ``classical'' (or, I would really add---although I don't think I *said* this---any form that one can see as suggested by Leibniz himself)'' (Stein, 2015, Personal Correspondence).}  On Stein's version, the argument concerns only the form that the dynamics would need to take, which could not involve ordinary differential equations governing the position and velocity of particles, since these give rise to precisely the sorts of trajectories that Stein argues are not invariant under automorphisms of Leibnizian space-times.  What is \emph{not} ruled out, for Stein, is a dynamics that would determine, from suitable initial conditions, (merely) relative positions and velocities for all subsequent times, since these \emph{would} be invariant under the sorts of transformations that preserve Leibnizian space-time.  Indeed, it is hard to see how one could want more than this out of a Leibnizian dynamics, since Leibniz apparently held that all motion was relative.  The difficulty, then, is how to express such a dynamics, i.e., to express a dynamics using only the resources of Leibnizian space-time.\footnote{It is worth noting that Stein's paper was delivered in 1975, whereas it was in 1977 that Barbour and Bertotti presented a dynamical theory arguably of the sort that Stein claimed had not been developed \citep{Barbour+Bertotti,Barbour+Bertotti2}.  (That said, as \citet{BelotBB} has emphasized, Barbour and Bertotti's theory cannot fully recover classical Newtonian gravitation, because it cannot deal with global rotation, and so whether it should count as an ``adequate'' Leibnizian dynamics is controversial.)  See \citet{BelotGM} for a detailed discussion of how one might express a suitably relationist dynamics, including the possibility of using a ``reduced'' or ``relative'' phase space, rather than space-time, as the starting point.}

\section{Two Concepts of Determinism}\label{sec:det}

The discussion above shows two things.  First is that for Earman, during the period when the 1987 hole argument paper was drafted, there appears to have been a close connection between the hole argument and Stein's argument; and second is that Earman's understanding of Stein's argument was importantly different from Stein's understanding of the argument.  The remainder of this paper will be devoted to diagnosing these differences, and then exploring how they bear on the interpretation and significance of the hole argument.

In the first instance, it seems that the difference concerns what should count as a deterministic theory.\footnote{Here there is a connection to a thread in the hole argument literature on what is sometimes called ``sophisticated determinism'' (as a contrast to ``sophisticated substantivalism''): see \citet{Melia}, \citet{BrighouseSD}, and \citet{NortonSD}.}  To get at the difference, consider first what sorts of assertions can be expressed using the resources of Leibnizian space-time.  Leibnizian space-time, by design, has the structure necessary to express: (1) assertions about duration between events; (2) assertions about instantaneous distance between bodies; (3) assertions about instantaneous angles between three or more bodies; and (4) assertions about how angles and distances change over time.  Together, these constitute facts about instantaneous configuration and relative motion.  These sorts of facts depend only on the Leibnizian structure $(M,t_a,h^{ab})$, and they are preserved under all isomorphisms---including automorphisms---of this structure.

As I read Stein, a deterministic Leibnizian dynamics would be one that could specify, uniquely, changes in these facts---that is, facts about instantaneous distances and angle---over time.  In other words, given an initial configuration, referring only to the sorts of facts that can be expressed using Leibnizian space-times as just described, a deterministic Leibnizian dynamics would specify subsequent facts of precisely (and only) that sort.  Stein's argument, then, concerns the resources available to express these dynamics.  Leibnizian space-time \emph{has} the resources to express these facts, including their change over time.  But standard ways of writing down dynamics require more than this: for instance, to express Newton's second law, $F=ma$, one needs to be able to make assertions about the acceleration of a body, not as a relative matter, but absolutely.  One cannot do this using only the structure of Leibnizian space-time, which can be seen by observing that the transformations Stein considers map curves to other curves with different acceleration (by any standard).  More generally, standard dynamics invoke differential equations, which generally require further structure, such as a derivative operator, to express.  Leibnizian dynamics, presumably, could concern only the dynamics of distance and angles between bodies, and not instantaneous rates of change of, for instance, velocity.

To get at this point in a different way, observe that one can always express differential equations by choosing a coordinate system and working with coordinate derivatives. But these dynamics---and their solutions---are not invariant under generic changes of coordinate system, much less arbitrary Leibnizian transformations.  And this is precisely the point.  Compare this situation to the one in Galilean space-time, where one \emph{does} have a preferred derivative operator, and so one can express differential equations whose solutions are invariant under arbitrary Galilean transformations, in the sense that any map that preserves the space-time structure will also map a solution to a solution---and, in fact, a solution that is the same as the original one with regard to all the assertions that can be expressed using the resources of Galilean space-time (which include, in addition to those of Leibnizian space-time, assertions about the acceleration of a body, and thus Newton's first and second laws).\footnote{These issues, regarding how to express dynamics in different space-time settings, have recently been discussed in detail in the context of Maxwellian space-time, where one has a standard of rotation but not acceleration. \citet{SaundersRNP}, for instance, argues that one can express suitable dynamics in this context using difference equations, rather than differential equations, whereas \citet{Dewar}, \citet{WeatherallComment}, and \citet{Chen} discuss more geometrical characterizations of (gravitational) dynamics in this setting.  See also \citet{WallaceWACS}.}

We can summarize these considerations compactly as follows.  On Stein's view, a deterministic Leibnizian dynamics would determine future states up to isomorphisms of Leibnizian space-time.  This is because to determine future states up to isomorphism is precisely to determine only those facts that are expressible within Leibnizian space-time.

Earman, by contrast, appears to demand more than this of a (deterministic) Leibnizian dynamics.\footnote{At least, he takes what follows to be a natural prima facie demand; one could read him as allowing that on reflection one might abrogate it.  I am grateful to an anonymous referee for emphasizing this point.}  He seems to require that such a dynamics determine, at least, unique worldlines for particles.  In other words, two histories, agreeing on the distances and angles between all particles over time, would be distinct if they disagreed on the locations occupied by the bodies over time, where ``location'' is characterized by the points occupied by the bodies' worldline/worldtubes.  This is strictly stronger than what Stein requires: location in this sense is not invariant under automorphisms of Leibnizian space-time, and so Earman's notion of determinism would require a Leibnizian dynamics to determine future states in a way that distinguishes between isomorphic possibilities.

Something subtle is going on here.  Earman suggests, in multiple places, that there are two different ways to respond to the argument Stein presents.  One of them, which Earman describes as the Leibnizian response, is to claim that the different models considered in Stein's argument represent the same possible history \citep[cf. ][p. 233]{Earman1986}.  The other response, which Earman suggests is Newton's strategy, is to ``beef up'' the structure of Leibnizian space-time, by adding further structure.\footnote{The availability of this second strategy in the classical case leads Earman to conclude that Stein's argument is in fact weaker than the hole argument \citep[p. 182]{EarmanWEST}; see also \citet[p. 233]{Earman1986}.}  I will have much more to say about how Earman understands the first of these responses in the next section.  But the second response is important here.

Earman seems to take for granted that Stein's argument fails for any space-time with at least the structure of Galilean space-time.  (Recall, above, his remark that Stein's argument shows that one requires at least as much structure as Galilean space-time \citep[p. 17]{Earman1989}; see also \citep[p. 233]{Earman1986} and \citep[p. 182]{EarmanWEST}.) But there is a certain sense in which even in Galilean space-time, no dynamics can uniquely determine the future locations of particles.  This follows from the hole argument itself, properly restated: choose, in Galilean space-time $(M,t_a,h^{ab},\nabla)$, some spacelike hypersurface $\Sigma$ (on which one will specify initial data), and a set $O\subset M$ with compact closure to the future of $\Sigma$.\footnote{We are taking for granted, here, that the space-time is temporally oriented.}  Now consider a diffeomorphism $d:M\rightarrow M$ that is the identity (only) outside of $O$.  Then $(M,d_*(t_a),d_*(h^{ab}),d_*(\nabla))$ is, again, Galilean space-time.

Thus, even if one specifies a dynamics using all and only the resources of Galilean space-time, and it determines some system of worldlines in Galilean space-time from initial data on $\Sigma$, the dynamics will not be sufficient to determine the future locations of particles.  This is because it cannot distinguish between a collection of trajectories $\{\gamma_i\}$ and trajectories $\{\gamma_i\circ d\}$, even though these both agree on $\Sigma$ and if one is a solution to the specified dynamics in Galilean space-time, so is the other.\footnote{I want to be clear that I do not take this argument to have any force---my point is that \emph{prima facie}, it should have force for Earman, given his other commitments.  (For further discussion of precisely what those commitments are, see the next section.)}

How is this argument different from Stein's?  It seems the only difference is that in Stein's argument, the diffeomorphism $d$ is an \emph{automorphism} of Leibnizian space-time---that is, $d$ is chosen so that $d_*(t_a)=t_a$ and $d_*(h^{ab})=h^{ab}$---whereas in the Galilean space-time argument I have just given, $d$ is an automorphism of the underlying manifold, but \emph{not} and automorphism of the Galilean space-time structure.  Instead, it determines an isomorphism of Galilean space-time that happens to be the identity map outside of a certain region.  The key, it seems, is that any automorphism of Galilean space-time that acts as the identity on certain initial data (in this case, $\Sigma$ and 4-velocities associated with points on $\Sigma$) necessarily preserves all timelike curves, whereas this is not true of Leibnizian space-time.

Why should this difference matter?  Earman does not say, and it is not perfectly clear why automorphisms should have a special status to him.\footnote{I do not want to put too much weight on this issue, because in some places Earman seems to suggest that the difference does \emph{not} matter.  For instance, \citet[p. 28]{EarmanPD} presents a schematized version of the hole argument, similar to what I have presented here for Galilean space-time, and suggests it gets at the core issue for Leibniz just as well as Stein's argument.   I will note, though, that he does not seem to connect this argument to determinism; and he also seems to think that the issue at stake is a certain failure of our modes of presentation of space-time models.  See also the next section.}  (For Stein, the difference matters because he is describing what would be required of a deterministic dynamics, and the automorphisms of each structure reflect what sorts of dynamics can be expressed using that structure.)  The idea seems to be that one fixes some fields as representing ``space-time structure'', and if one can come up with a transformation that preserves initial data and preserves all of the space-time structure, but does not preserve particle worldlines, then a theory in that space-time setting is necessarily indeterministic.

I will not attempt to recover what Earman had in mind in further detail.  It suffices to point out a certain consequence of his view, which is: Earman apparently requires a deterministic Leibnizian dynamics to specify future facts of a sort that Leibnizian space-time does not have the resources to express---and of a sort that Leibniz would have disavowed.  In other words, Earman's view renders Leibnizian space-time an unacceptable setting for recovering Leibniz's views, because it is not a suitable structure for a theory of how relative positions evolve over time.

\section{Manifolds, Substantivalism, and Structure}\label{sec:sub}

I noted in the previous section that Earman described two ways to respond to Stein's argument.  One, which I have already discussed, involved ``beefing up'' Leibnizian space-time structure.  The other was to insist that two models, related by the Steinian transformation, characterize the same physical possibilities.  I will now discuss the second of these, which Earman calls the ``Leibnizian response''.

One way of understanding Earman's Leibnizian response would be as the view I have already attributed to Stein: that is, to say that Leibnizian space-time has the resources to express facts about relative position and angle, over time; and thus to suppose that it is only those facts, represented by worldlines or worldtubes in Leibnizian space-time, that are to be attributed to the world.  On this view, any two isomorphic Leibnizian space-times (with worldlines and worldtubes appropriately transformed along the isomorphism) represent the same physical situations, because, by construction, they agree on precisely the facts that are taken to be physically significant.  Indeed, this is automatic and nearly trivial, since isomorphic models agree on precisely those assertions that can be expressed in Leibnizian space-time---which are just those assertions that one has adopted Leibnizian space-time to represent.  Leibnizian space-time is just a way of characterizing certain relative facts about changes in instantaneous configuration of bodies: no more, and no less.

Of course, this would \emph{not} be a response to Stein's argument as Stein understood it, because Stein already took this attitude for granted in his argument;\footnote{Curiously, in \citep[note 4]{Earman1989}, Earman says Stein is ``perfectly aware that this is the kind of response that the relationist would make, but he is skeptical that the response can be made to stick.''  On my reading of Stein, this response is not to the point, much less the one Stein anticipates the relationist making.  I cannot find anywhere in the paper where Stein suggests that such a response cannot ``be made to stick''.} to respond to Stein's argument one would need to, in addition, provide a dynamics for relative configurations, using only the resources of Leibnizian space-time.  It is also, I think, not quite how Earman understood the situation.  The reason is that Earman believed that to adopt the Leibnizian response, one would need to make an adjustment to Leibnizian space-time---one that amounts to rejecting ``substantivalism''.

What could this mean?  Consider, first, how Earman characterizes the ``Leibnizian'' response.  For instance, in \citep{Earman1989}, he writes:
\begin{quote}\singlespacing ...it is open to the relationist to reply that the violation of determinism is an illusion fostered by the false presumption that space-time is a substance; for on a non-substantivalist conception of space-time, the original solution and its image under the symmetry map are merely different modes of presentation of the same physical history. (p. 13)\end{quote}
Likewise, in \citep{Earman1986}, he writes:
\begin{quote}\singlespacing
The failure of determinism here is only an apparent one, the false appearance being due to the chimerical assumption of the reality of space-time in itself.  Dropping that assumption and realizing that space-time is something merely relative, we see that [the Steinian argument yields] different ways of representing the same physical history. (p. 233) \end{quote}
What is striking in these passages is the idea that somehow, in running Stein's argument, one has implicitly assumed substantivalism.  And yet, I think it is clear that \emph{Stein} did not take the argument to assume substantivalism---since, again, doing so would vitiate any claim to recovering Leibniz's views.

The key to resolving the mystery is that Earman seems to have taken \emph{any} space-time structure in which a manifold is used to represent events in space and time to, \emph{ipso facto}, carry with it substantivalist commitments.  He says as much explicitly in \citep{Earman1986}, writing:
\begin{quote}\singlespacing I began with space-time models $M, O_1, O_2,\ldots$, which, \emph{prima facie}, seem to call for a full-blown substantivalism in which the space-time manifold $M$ is the only substance or basic object of predication and the fields $O_i$ are properties of $M$.  Then in order to avoid an unpleasant consequence, I concluded that these models must be grouped into equivalence classes, where $M, O_1, O_2,\ldots$ and $M', O_1',O_2',\ldots$ are counted as equivalent just in case there is a diffeomorphism $d:M\rightarrow M'$ (onto) such that $d^*O_i = O_i'$ for each $i$, and that an equivalence class corresponds to a single physical reality with different members of the class being merely different representations of that reality. (pp. 236-7)\end{quote}
Likewise, when he defines space-time structures in \citep{Earman1989}, he calls models of the form we have been considering, and of the form one encounters throughout relativity and contemporary philosophy of physics, \emph{substantival} models:
\begin{quote}\singlespacing A \emph{substantival space-time model} has the form $M, O_1, O_2,\ldots$ where $M$ is the space-time manifold (the collection of space-time points equipped with a differentiable structure) and the $O_i$ are geometric object fields on $M$. All familiar space-time theories, whether classical, special relativistic, or general relativistic, can be formulated in this form. (p. 17, emphasis original)\end{quote}
So it seems that what Earman has in mind as the ``Leibnizian'' response is not only to recognize that Leibnizian space-time only has the resources to express facts about relative motion---but also, to demand that one adopt some formulation of Leibnizian space-time in which a manifold does not appear, lest one saddle the Leibnizian with substantivalism.  This, then, is what leads Earman to propose Leibniz algebras as a candidate alternative to Leibnizian space-time: they avoid the implicit commitment to substantivalism that Earman believes comes along with space-time structures that use manifolds.

There are several remarks to make at this point.  The first is just to emphasize again that Earman here takes space-time structures of the form $(M,O_1,O_2,\ldots)$ to have expressive resources---roughly, ``structure''---that is not preserved by diffeomorphisms.\footnote{An anonymous referee raises the concern that speaking of ``structure preserved by diffeomorphisms'' is Procrustean.  It is true that Earman did not express himself in this way.  But these ideas were widespread in the 1970s and 1980s, when questions of definability were widely discussed within philosophy of physics. So it is not anachronistic, even if it is not Earman's idiom.  On the other hand, one might contend that it is a substantive view in the philosophy of applied mathematics or philosophy of scientific representation---one that Earman might well reject---that looking at what is invariant under certain classes of transformation should be the primary route to identifying the representational capacities of mathematical objects.  But from this perspective, which I think is apt, my point is just that Earman apparently \emph{did} reject this view.}  This is so because he takes the failure of determinism given by the Steinian argument to be a failure to determine the subsequent locations of---that is, points of $M$ occupied by---bodies in Leibnizian space-time.  But by Stein's argument or the modified hole argument described above, this is precisely the sort of thing that is \emph{not} preserved by diffeomorphisms, even when one pushes forward the rest of the space-time structure.  And yet, Earman does not conclude from such arguments that these sorts of facts cannot be expressed using the resources of Leibnizian space-time (since they are not invariant under its automorphisms); instead, he apparently concludes that they \emph{can} be so expressed, but no dynamics can determine them uniquely.  Given the character of Stein's argument as I have reconstructed it, this conclusion is sharply at odds with how Stein understood the situation: after all, the point of the argument was to emphasize what sorts of assertions could be expressed using the resources of Leibnizian space-time, by pointing out what is and is not preserved under its automorphisms.  The ``substantivalism'' that Earman takes to be implicit in the structure amounts to the ability to express assertions of a sort that are not invariant in this way---just as statements of the form ``this particle is unaccelerated'' is not expressible in Leibnizian space-time.

The second remark is that there is a certain irony to all of this.  In \citep{EarmanWEST}, and indeed, throughout his work on substantivalism and relationism during the 1970s and 1980s, Earman carefully distinguishes between ``relationist'' conceptions of space-time and ``relativist'' conceptions of motion, noting that neither implies the other (and, similarly, that their ``subtantivalist'' and ``absolutist'' alternatives, respectively, also do not imply one another).  This distinction is important, correct, and seems not to have been clearly made previously.\footnote{That said, one can identify a similar distinction in Stein's suggestion, for instance in \citep{SteinNST}, that substantivalism was not particularly important to Newton (especially given that Newton explicitly disavowed it in \emph{De Gravitatione...}); but that the doctrine of absolute space and time was essential to Newton's understanding of his laws of motion.  On this view, it was arguably Leibniz, and perhaps Clark, who mixed the two views, producing a morass that took centuries to extract ourselves from.}   But Earman's argument for this distinction seems to rely on his insistence that any characterization of space-time structure that begins with, or includes, a manifold of space-time events is necessarily substantivalist.  The irony is that these sorts of models were introduced to philosophers, at least in the context of classical physics, by \citet{SteinNST}, whose express purposes was to make a point about motion and laws, \emph{without} endorsing substantivalism.  Another irony, I suppose, is that space-time models of the sort Earman considers emerged out of work by Einstein, who likewise explicitly disavowed a substantivalist interpretation: as Einstein writes, and as Earman often quotes: ``Space-time does not claim an existence of its own but only as a structural quality of the [metric] field'' \citep[p. 155]{Einstein1961}.\footnote{See, e.g., \citet[p. 17]{Earman1989}.  Einstein's book was originally published in December 1916, but the quote in question was only added to the 15th edition, as part of a new Appendix.}  It is hard to see why one should be compelled to take these models as somehow implying substantivalist commitments.

So why \emph{does} Earman take space-time models of this form to be inherently ``substantivalist''?  One plausible answer runs through Quinean ontology.\footnote{I am grateful to Jeremy Butterfield for suggesting this diagnosis---one he also applied to himself \citep{Butterfield}---though I am not sure if he endorses it.  \citet{BelotDiss} also appears to link Quine's dictum and a general commitment to realism to Earman's understanding of substantivalism.  On the other hand, I do not have explicit textual evidence of Earman endorsing Quine's view, and so Earman's contention that use of a manifold is metaphysically loaded may well have a different origin.  }  Recall Quine's oft-quoted (and sometimes mocked) dictum: ``To be is, purely and simply, to be the value of a variable'' \citep[p. 32]{QuineOWTI}.\footnote{In more careful moments, Quine expressed a subtly different view: e.g., ``The ontology to which an (interpreted) theory is committed comprises all and only the objects over which the bound variables of the theory have to be construed as ranging in order that the statements affirmed in the theory be true'' \citep[p. 11]{QuineOI}.  Here, I take it, one must do more than merely peruse the quantifiers of one's physics textbooks to determine one's ontological commitments.}  If one adopts such a view, then flipping through classic texts on general relativity reveals physicists quantifying over points $p$ in $M$ all over the place.  And so, it seems, Quinean physicists must be committed to the existence of space-time points, in some sort of ontologically robust ``realist'' sense of ``existence''.  More generally---leaving Quine aside---Earman takes it that the manifold (or its points) appears to be the ``basic object of predication'' in these structures.\footnote{This argument echoes \citet{Field}, who argues for substantivalism on the grounds that without it one cannot make sense of fields as distributions of properties (because those properties are, apparently, properties of space-time regions).  \citet[note 20]{Earman1986} describes Field's arguments as ``the most interesting and sustained defense of space-time substantivalism to be found in the modern philosophical literature''.}  In other words, it seems that there is an immediate realist reading of any theory whose models include a manifold as one on which that manifold reflects putative facts about the locations of events.

\section{The Hole Argument, Revisited}\label{sec:holes}

I will conclude by returning to the hole argument in light of the considerations raised above.  First, as I argued in section \ref{sec:det}, Stein and Earman are working with two different views of what it would take to have a deterministic theory in the context of Leibnizian space-time. Now compare these views of determinism, translated to general relativity.  Given initial data for Einstein's equation, the Steinian view would be that a deterministic relativistic dynamics should determine the evolution of the metric up to isomorphism (isometry); whereas the Earmanian view would be that a deterministic relativistic dynamics must determine the metric uniquely at each space-time point---unless one modified the formalism in some way.

How different are these?  Of course in the end, Earman concludes that we ought to adopt the Steinian view.\footnote{In this sense, then, calling the second view the ``Earmanian view'' is unfair---though it does reflect what Earman takes to be the ``default'' position.}  What matters, though, is the route to this conclusion.  In the Leibnizian case, Stein, as we have seen, takes the argument to start from a structure with certain expressive resources, and draws a conclusion about what sort of dynamics one can have.  Earman, meanwhile, takes the argument to start from a structure that presupposes ``substantivalism'' and concludes that substantivalism is incompatible with determinism (of any sort) in Leibnizian space-time---and thus, to avoid the dismal conclusion, one must modify Leibnizian space-time by moving to some other structure, such as Leibniz algebras.

This latter version of the argument looks strikingly similar to the starting and ending points of the hole argument, as presented, for instance, in \citep[pp. 234--9]{Earman1986}, \citep[Ch. 9]{EarmanWEST}, or even \citep{Earman+Norton}.  There, too, one begins with a structure $(M,O_1,O_2,\ldots)$ that is taken to presuppose ``substantivalism'' and, by considering the action of diffeomorphisms from $M$ to itself, one concludes that substantivalism is incompatible with determinism (of any sort) in a relativistic space-time---unless one modifies the formalism of general relativity.  In \citep{Earman1986} and \citep{EarmanWEST}, Earman goes on to propose moving to Einstein algebras as a suitable modification, though he also takes it that ``forming equivalence classes'' under isometry would capture the same idea.  The important point is that Earman understands the hole argument just as he understood the Steinian argument: it shows that there is an inadequacy in the standard formalism of general relativity that arises from the apparent substantivalist commitments of that formalism.  It is an argument that Earman takes to reveal that general relativity exhibits ``excess structure''.\footnote{In fact, Earman refers to the ``space-time points'' of the manifold as ``descriptive fluff'' \citep[p. 170]{EarmanWEST}---an expression he famously repeats in his oft-quoted critique of textbook presentations of the Higgs mechanism: ``neither mass nor any other genuine attribute can be gained by eating descriptive fluff'' \citep[p. 190]{EarmanCurie}.  This suggests an understanding of the manifold as strongly analogous to ``gauge structure'', as in electromagnetism.  See \citet{WeatherallUG} and \citet{Bradley+etal} for discussions of the sense in which this sort of ``structure'' is ``excess'', along with further arguments that general relativity does not have excess structure in the sense that some formulations of electromagnetism do.}

Stein never published his reaction to the hole argument.  But by studying his understanding of his argument in Leibnizian space-time, one can extract a different view of what the hole argument accomplishes.\footnote{Some (weak) evidence that Stein would not think of the hole argument as bearing on ``substantivalism'': ``You [Gr\"umbaum] tend to think of the world in terms of ``things''---``primary substances,'' in Aristotle's sense.  I do not: I tend, rather, to think ... of ``structures'' and ``aspects of structure'' \citep[395]{SteinGrunbaum}.}  It begins with a structure with certain expressive resources---those of a Lorentz-signature metric on a manifold---and it concludes that the only facts that can be determined by a dynamics set in that structure are those that are preserved by isometry.  This payoff is somewhat different from in the Leibnizian space-time case, because one \emph{can} express partial differential equations, using the covariant derivative operator determined by the metric, in a relativistic space-time.  But it does show that the notion of ``initial value formulation'' one should expect to get in general relativity will be somewhat different from classical treatments of symmetric hyperbolic systems of partial differential equations: in particular, we can expect unique solutions only up to isometry.  This is an important, even deep, conclusion about the character of differential equations in general relativity---and one that arguably has still not been fully understood.  But it is a different moral from the one Earman draws.

So it seems we have two different ways of understanding the hole argument, just as with Stein's argument. One is about substantivalism and excess structure; the other is about the character of dynamics and differential equations.  That these two understandings are distinct, and both were present in the literature, in some form, in the 1970s, is the principal moral I hope to draw from this discussion.

That said, it is worthwhile to reflect on which of these understandings is to be preferred.  I think Earman's understanding has dominated the subsequent literature.  But there are several considerations to recommend what I have reconstructed as a ``Steinian'' attitude towards the hole argument (with some unfairness, as Stein never expressed this view).  One is that Earman's proposal for how to amend the formalism of general relativity in response to the hole argument, by introducing Einstein algebras, signals a problem with the idea that the hole argument reveals that the standard formalism has ``excess structure''.  This is because there is a strong sense in which Einstein algebras and relativistic space-times are equivalent with regard to how much structure they have: the category of relativistic space-times is dual to the category of Einstein algebras \citep{Rosenstock+etal}.\footnote{For discussions of this criterion of equivalence of structure, see \citet{BarrettStructure} and \citet{WeatherallTE,WeatherallUG,WeatherallReview,WeatherallWNCE}.}   This result captures the sense in which every Einstein algebra is associated with a relativistic space-time, unique up to isometry, and vice versa; but it also captures a sense in which every isometry of relativistic space-times, including those that are used in the hole argument, corresponds uniquely to an isomorphism of Einstein algebras.

This latter fact, which Earman himself seems to have recognized (though perhaps not fully appreciated),   was the basis of a classic argument by \citet{Rynasiewicz1992} that Einstein algebras could not accomplish what Earman hoped.  The reason, however, is not that Einstein algebras, too, have excess structure.\footnote{For Earman's reservations on this point, see \citet[p. 239]{Earman1986} and \citet[p. 193]{EarmanWEST}, where he observes that one gets a proliferation of isomorphic Einstein algebras corresponding to diffeomorphic manifolds.  He takes this to show that Einstein algebras may remove substantivalist commitments from space-time models, but only to the ``first degree''; there is some higher-order sense, which he does not elaborate, in which Einstein algebras should still be seen as substantivalist.  What \citet{Rynasiewicz1992} adds to this are the twin observations that one can recover ``points'' from Einstein algebras as certain maximal ideals; and that the diffeomorphisms used in the hole argument give rise to isomorphism of Einstein algebras whose action on these ideals corresponds to the action of the diffeomorphisms on points.}  Rather, it is that the ``excess structure'' of relativistic space-times was a chimera, arising because one took manifolds to include structure that is not preserved by diffeomorphism.  To think otherwise is to mistake the structure that mathematicians intend by ``manifold''---a structure that is explicitly defined only up to diffeomorphism---for something with further expressive resources.\footnote{As an aside, this perspective makes sense of Einstein's claim about manifolds being structural qualities of fields: a manifold is ``nothing but'' a means of characterizing a space of possible field configurations.  It also provides a hint at how to solve Field's dilemma about fields \citep{Field}.}

As I argue in \citet{WeatherallHoleArg}, that manifolds do not have these resources is not by itself an argument against substantivalism.  After all, one could introduce a new structure, above and beyond manifolds, that would permit one to represent the further facts about where events occur that the substantivalist wants to accept.  Following \citet{StachelSS}, one possible structure would be a manifold endowed with a smooth individuating field---essentially, a field of ``labels'' distinguishing points.  The fact that one need not invoke such structure in working with general relativity does not imply that it is not instantiated by the world.  But the crucial point is that in order to express the substantivalist's facts, one must \emph{add} structure to the standard formalism of general relativity, which stands in marked contrast to Earman's conclusion that the hole argument reveals that one must \emph{remove} structure from the standard formalism in order to avoid commitment to these substantival facts.

From this perspective, the hole argument can be recast as, once again, an argument against substantivalism.  This is because the hole argument shows that the person who endorses this ``label'' structure as representing true but unobservable facts about the world apparently \emph{is} committed to a certain kind of indeterminism.\footnote{Though one should be careful: as \citet{NortonSD} emphasizes, what the hole argument shows is only that Einstein's equation---or more generally, a dynamics using only the structure of a Lorentzian manifold---cannot determine these further facts.  But one should not have expected Einstein's equation to do so, and the fact that Einstein's equation cannot determine these facts does not mean that these facts are not determined by other considerations (or even other laws).}  This might be good reason to reject the substantivalist's label structure.  But doing so does not necessitate moving to some new, weaker structure for representing space-time; to the contrary, it recommends a retreat to precisely what we started  with.

\section*{Acknowledgments}
This material is partially based upon work produced for the project ``New Directions in Philosophy of Cosmology'', funded by the John Templeton Foundation under grant number 61048. I am grateful to John Earman, David Malament, John Norton, Howard Stein, and two anonymous referees for insightful conversations in connection with this paper (though I have no reason to believe any of them endorse my arguments), and to David Malament and Howard Stein for comments on a previous draft.  Earlier versions of this paper were presented in Varna, Bulgaria and London, UK; I am grateful to the organizers of those meetings and to the participants for helpful discussions.  I am also grateful to the participants in my graduate seminar on the philosophy of Howard Stein, held in Winter 2016, for discussion of this material.

\singlespacing

\bibliographystyle{plainnat}
\bibliography{holes}

\end{document}